# Machine Learning Operations: A Mapping Study

Abhijit Chakraborty
School of Computing and Augmented Intelligence
Arizona State University
achakr40@asu.edu

Suddhasvatta Das
School of Computing and Augmented Intelligence
Arizona State University
sdas76@asu.edu

Kevin Gary
School of Computing and Augmented Intelligence
Arizona State University
kgary@asu.edu

***Abstract*** **- Machine learning and AI have been recently embraced by many companies. Machine Learning Operations, (MLOps), refers to the use of continuous software engineering processes, such as DevOps, in the deployment of machine learning models to production. Nevertheless, not all machine learning initiatives successfully transition to the production stage owing to the multitude of intricate factors involved. This article discusses the issues that exist in several components of the MLOps pipeline, namely the data manipulation pipeline, model building pipeline, and deployment pipeline. A systematic mapping study is performed to identify the challenges that arise in the MLOps system categorized by different focus areas. Using this data, realistic and applicable recommendations are offered for tools or solutions that can be used for their implementation. The main value of this work is it maps distinctive challenges in MLOps along with the recommended solutions outlined in our study. These guidelines are not specific to any particular tool and are applicable to both research and industrial settings.**

## I. Introduction

Machine Learning (ML) has become an important process to leverage the potential of data and allows businesses to be more innovative [1], efficient [2], and sustainable [3]. It has emerged alongside big data technologies and high-performance computing, creating new opportunities to analyze and understand data-intensive processes in various operational environments. The learning process in machine learning involves the ability of machines to learn from experience and perform tasks without being strictly programmed. However, as highlighted in [4], managing and maintaining machine learning systems poses unique challenges compared to traditional software systems. The success of many productive ML applications in real-world settings falls short of expectations [5]. Many ML projects fail—with many ML proofs of concept never progressing as far as production [6]. From a research perspective, this does not come as a surprise as the ML community has focused extensively on the building of ML models, but not on (a) building production-ready ML products and (b) providing the necessary coordination of the resulting, often complex ML system components and infrastructure, including the roles required to automate and operate an ML system in a real-world setting [7]. For instance, in many industrial applications, data scientists still manage ML workflows manually to a great extent, resulting in many issues during the operations of the respective ML solution [8]. As the field of artificial intelligence and machine learning continues to evolve, the need for efficient and reliable deployment of these technologies has become increasingly crucial.

The adoption of DevOps principles in software engineering has enabled developers to deliver their products more efficiently and scalable, and this trend is now being applied to machine learning projects, resulting in the emergence of MLOps. Machine Learning Operations, or MLOps, has emerged as a critical discipline that aims to address the unique challenges associated with running and maintaining machine learning systems in production environments.

### A. DevOps for ML Systems - MLOps

DevOps is a subset of software engineering focused on tightening the coupling between the development and operation of software systems. DevOps principles advocate for end-to-end automation [17] which is expressed through the use of version control systems, automated build and deploy pipelines, etc. Some motivating factors for automation are shortening the time to delivery, increasing reproducibility, and reducing time spent on automatable processes [18]. DevOps for Machine Learning, named MLOps, is a subset of SE for ML and a superset/extension of DevOps, focusing on adopting DevOps practices when developing and operating ML systems [19]. [20] defines that "ML Ops is a cross-functional, collaborative, continuous process that focuses on operationalizing data science by managing statistical, data science, and machine learning models as reusable, highly available software artifacts, via a repeatable deployment process." [20] identifies four main steps of MLOps: Build, Manage, Deploy and Integrate, and Monitor. Fig. 1 shows the MLOps pipeline proposal in [21]

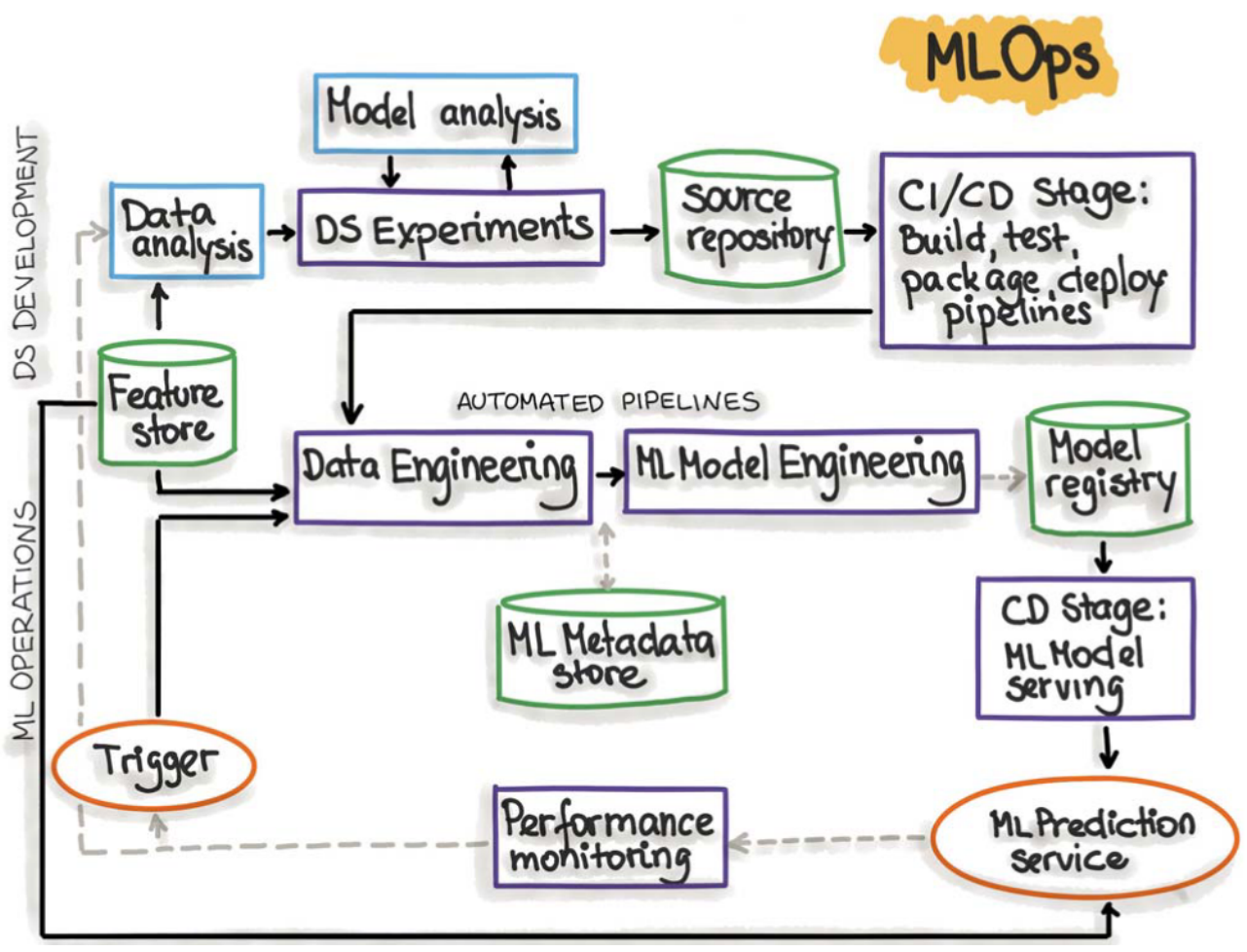

Fig 1. MLOps Pipeline [21]

### B. MLOps

MLOps is the collection of techniques and tools for the deployment of ML models in production [35]. Encompassing the combination of DevOps and Machine Learning processes,

DevOps [36] represents the set of practices minimizing the needed time for a software release, reducing the gap between software development and operations [37] [38]. The two main principles of DevOps are Continuous Integration (CI) and Continuous Delivery (CD). Continuous integration is the practice by which software development organizations try to integrate code written by developer teams at frequent intervals. They constantly test their code and make small improvements each time based on the errors and weaknesses that results from the tests. This results in a reduction in the software development process cycle [39]. Continuous delivery is the practice according to which, there is constantly a new version of the software under development to be installed for testing, evaluation and then production. With this practice, the software releases resulting from the continuous integration with the improvements and the new features reach the end users much faster [40]. After the great acceptance of DevOps and the practices of "continuous software development" in general [41] [36], the need to apply the same principles that govern DevOps in machine learning models became imperative [12].

MLOps attempts to automate Machine Learning processes using DevOps practices and approaches, seeking the rapid automated delivery benefits of DevOps. MLOps specifically attempts to apply CI and CD principles within the ML model development, integration, and deployment phases [37]. Although it seems straightforward in reality it is not. This is due to the fact that a Machine Learning model is not independent but is part of a wider software system and consists not only of code but also of data. As the data is constantly changing, the model is constantly called upon to retrain from the new data that emerges. For this reason, MLOps introduce a new practice, in addition to CI and CD, that of Continuous Training (CT), which aims to automatically retrain the model where needed. From the above, it becomes clear that compared to DevOps, MLOps are much more complex and incorporate additional procedures involving data and models [45] [33] [42].

One of the key challenges in MLOps is the operationalization of the complex lifecycle of machine learning models in production. Unlike traditional software systems, which are typically governed by a well-defined set of instructions, machine learning models are constantly evolving and require careful monitoring and maintenance to ensure their continued performance and reliability [16]. As Sculley et al. [12] noted, "the long-term costs in ML systems" can be significant, with issues such as managing the power configuration for multiple models, tracking experiment results, and monitoring the entire production pipeline.

Existing studies have shown that the operationalization of ML is an area that presents practitioners with real challenges [14]. Operationalization, in the context of this paper, consists of taking a trained and evaluated ML model to a serving state in the intended production environment, including necessary support functions, such as monitoring. Tackling operationalization challenges requires adopting good practices and utilizing suitable tooling or solutions.

Considering the intricate challenges and evolving nature of MLOps, embracing a research approach that can effectively explore the wide range and profound aspects of the area is essential. Given this, we have chosen to conduct a Systematic Mapping Study (SMS) as our research methodology. This choice is supported by numerous compelling arguments that are in line with our aims.

A MLOps system includes various pipelines [27]. Commonly a data manipulation pipeline (DM), a model creation pipeline (MC) and a deployment pipeline (MD) are mandatory. Each of these pipelines must be compatible with the others, in a way that optimizes flow and minimizes errors.

### *A. MLOps Pipelines*

MLOps (Machine Learning Operations) pipeline is a collective term used to describe a collection of three different pipelines that are essential for managing the end-to-end lifecycle of machine learning models, from development and training to deployment and monitoring. These pipelines help ensure that machine learning models are scalable, reliable, and maintainable in production environments. Below are descriptions of each of the pipelines in detail.

*Data Management Pipeline* (DM): Involves gathering, cleaning, and preprocessing the data needed to train and evaluate machine learning models. It may involve feature engineering to extract relevant information from raw data.

*Model Creation Pipeline* (MC): Machine learning models are trained on the prepared data. This includes selecting appropriate algorithms, tuning hyperparameters, and evaluating model performance using techniques like cross-validation. After training, models are evaluated using validation datasets to assess their performance metrics, such as accuracy, precision, recall, or F1 score. This step helps ensure that models generalize well to unseen data.

*Model Deployment Pipeline* (MD): Once a model meets the desired performance criteria, it is deployed to production environments where it can make predictions on new data. Deployment involves packaging the model into a deployable format, integrating it with existing systems, and setting up APIs for inference and monitoring for it's stability.

*Sustainability Pipeline* (Sustainability): Focusing on integration and sustainability considerations in the development, deployment, and maintenance of machine learning models. Sustainability in this context encompasses environmental, social, and economic aspects, aiming to minimize negative impacts and maximize positive outcomes throughout the ML pipeline lifecycle. This refers to the sustainable factors for each of the above three pipelines.

This study focuses on 1) researching ML operationalization in-depth, and not as part of a broader study of SE for ML and 2) putting more focus on tooling/solutions and infrastructure by identifying novelty in how current solutions are proposed to be used and what is reported to need further research. These objectives are fundamental for understanding the complex dynamics of MLOps and essential for guiding future academic and practical efforts. With this motivation, we framed the following research questions:

- **RQ1**: What are the research trends in MLOps, how many studies cover these and how do they converge to different MLOps pipelines?
- **RQ2**: What kind of novelty or new ideas do these studies constitute in relation to the pipelines?

## II. RELATED WORK

ML models are increasingly prevalent in virtually all fields of business and research in the past decade. [15] said that half of the organizations questioned had used artificial intelligence in some commercial operations. Despite extensive research on training and evaluating machine learning models, the primary challenge for many companies and practitioners lies not in discovering new algorithms and optimizations for training, but rather in effectively deploying models to production to generate tangible business value. Most companies are still in the very early stages of incorporating ML into their business processes [16].

Considering that MLOps is a relatively young topic, it is not surprising that there are not a lot of review publications. In this part, we will first present the review articles, and then we will discuss some of the most significant and impactful work that has been done in each and every job in the MLOps life cycle. Following the presentation of a basic overview of MLOps by Goyal [25], Zhao [26] examines the academic literature about machine learning in production in order to determine the significance of MLOps. In addition, Zhou et al. [27] focuses on the usage of resources over the whole life-cycle of MLOps. In addition to reviews, there are a great number of papers that discuss the applications of MLOps in a variety of fields. Some examples of these papers include the MLOps approach in the cloud-native data pipeline design by Poloskei [28], the application of MLOps in the prediction of lifestyle diseases by Reddy et al. [29], and SensiX++: Bringing MLOps and Multi-tenant Model Serving to Sensory Edge Devices by Min et al. [30].

In terms of the various phases of MLOps, Makineth et al. [31] highlight the significance of MLOps in the area of data science. Their findings are based on a survey in which they gathered answers from 331 experts hailing from 63 different countries. Regarding the data manipulation task, Renggli et al. [32] explain the relevance of data quality for an MLOps system while demonstrating how different characteristics of data quality propagate through various phases of machine learning development. This is done in relation to the data manipulation job. In the MLOps cycle, Ruf and colleagues [33] investigate the function that the MLOps tools play as well as the connection between them for each and every activity. Additionally, they provide a formula for selecting the best Open-Source tools that are currently available. Klaise and colleagues [34] had a discussion on monitoring and the issues that are associated with it. They used current examples of production-ready solutions that were developed utilizing open source tools. Finally, Tamburri [37] discusses the difficulties and tendencies that are now occurring, with a particular emphasis on explainability and sustainability.

The contribution of this paper would be two-fold to the greater book of knowledge, first would be an aggregated view of categorizing the research trends under different MLOps pipelines with novelty of usage of existing tools and solutions applications in each of these pipelines, that are being proposed by the existing research, this would help the readers to understand the potential trends. The second would be to identify the potential gaps that the community would need to address in future research to come up with effective strategies for operationalizing the MLOps pipelines.

## III. Methods

Systematic mapping studies (SMS), also known as scoping studies, aim to provide a comprehensive picture of a study topic by categorizing and quantifying the contributions within the categories of that area ([4],[9]). This approach entails doing a comprehensive analysis of the current body of literature in order to get a deep understanding of the topic range and the various forms of publishing. Scholars in many domains use systematic mapping research, which adheres to certain principles or techniques [4]. An SMS is especially suited for offering an exhaustive overview of the research environment, therefore accomplishing the identification of knowledge gaps that need additional research. Additionally, an SMS aligns with our goal to categorize current information and assemble guidelines that will impact future research endeavors. During this study, we followed the systematic mapping recommendations described by [9], specifically using this methodology in the field of software engineering.

Following the method of Webster & Watson [22], and Kitchenham et al. [23]The mapping study was carried out using a four-step process: 1) Identifying the research questions, 2) Performing an extensive search for relevant literature, 3) Choosing research of superior quality that matches the predetermined requirements 4) Analyzing and consolidating data from the chosen research to uncover recurring themes and patterns.

### A. Study Selection

*Library Scan:* The primary aim of this phase is to identify studies that are in line with the research topics previously stated. This section provides a comprehensive description of the selection process carried out by the lead investigator and later evaluated by the other authors. The year 2015 marked a pivotal moment in the evolution of Machine Learning Operations (MLOps), a crucial discipline that bridges the gap between the development and deployment of machine learning (ML) models [12]. Prior to 2015, the landscape of ML tooling and frameworks was fragmented, with organizations struggling to effectively manage the lifecycle of their machine learning systems [24]. However, the rising prominence of MLOps in 2015 signaled a transformative shift in the way organizations approached the challenges inherent in productionizing machine learning models. One of the key drivers behind the increased focus on MLOps in 2015 was the growing recognition of the unique challenges posed by deploying and maintaining machine learning systems in production environments [10].

First we constructed a brute-force search query on Google Scholar for "Machine learning Ops" that got more than 26,100 results. We decided to refine our search, and the subsequent string was "(Machine learning Operations) AND (Machine learning Ops OR MLOps)," and the number of results went down to 3,460. With further filtering based on ‘review articles’ the final number stood up to 269. Initially, our search was conducted on Google Scholar (which, by default, comprehensively covers the most significant databases). Subsequently, we applied semantically identical search strings to the other three databases, namely, ACM Digital Library, IEEE Xplore, and Scopus, which were searched with the strings. However, Google Scholar does not allow refined search strings to limit the searches to just abstracts/titles, but the other three databases provide this level of refinement. Thus, we utilized that to generate a focused list of studies. While Scopus and IEEE were giving relevant results, ACM was not providing anything, hence the search criteria was modified from ‘Title’ to ‘Anywhere’ in the document, and post application of the year filter it provided 289 articles. Lastly, given the number of studies retrieved from our database search, we categorized them based on relevance, limiting our exclusion pages in the databases, with each containing ten studies, where the relevance of the studies noticeably declined to a document to no relevant document for the pertaining page.

- **ACM**: [[All: "machine learning operations"] OR AND [[All: "machine learning operations"] OR [All: "mlops"]] AND [E-Publication Date: (01/01/2015 TO 12/31/2024)]]
- **IEEE**: "Document Title": ("Machine learning Operations" AND ("Machine learning Ops" OR "MLOps") ) AND "Publication Year": 2015 -2024
- **Scopus**: TITLE-ABS-KEY ( "Machine learning Operations" AND ( "Machine learning Ops " OR "MLOps" ) ) AND $PUBYEAR > 2015$ AND $PUBYEAR < 2024.
- **Google Scholar**: "Machine learning Operations" AND ("Machine learning Ops " OR "MLOps" ) AND (Year: 2015 – 2024) AND (Article Type: Review only)

*Title and Abstract:* All abstracts were read and if the paper did not present information on MLOps it was excluded. This reduced the number of studies to 81. This was done concurrently with the next step.

*Duplicates Removed:* The publications that are present in multiple databases are removed manually which comprised of 34 duplicate studies. This step helped us to find the unique individual work that came down to 49 studies from 81 studies obtained from the previous step before the execution of duplication removal.

*Full-Text Scan:* The final selections of papers were based on a thorough review of studies from the above step, carried out by the authors. We focused on the research questions pertaining to MLOps and the overall quality of the papers. Assessing paper quality can be subjective [13], so we looked at factors like where it was published, when it was published, how often it was cited (if it has not been published in recent years), and the reputation of the venue and authors. Those papers which are secondary studies based on the MLOps concepts, they were not considered as part of this study.

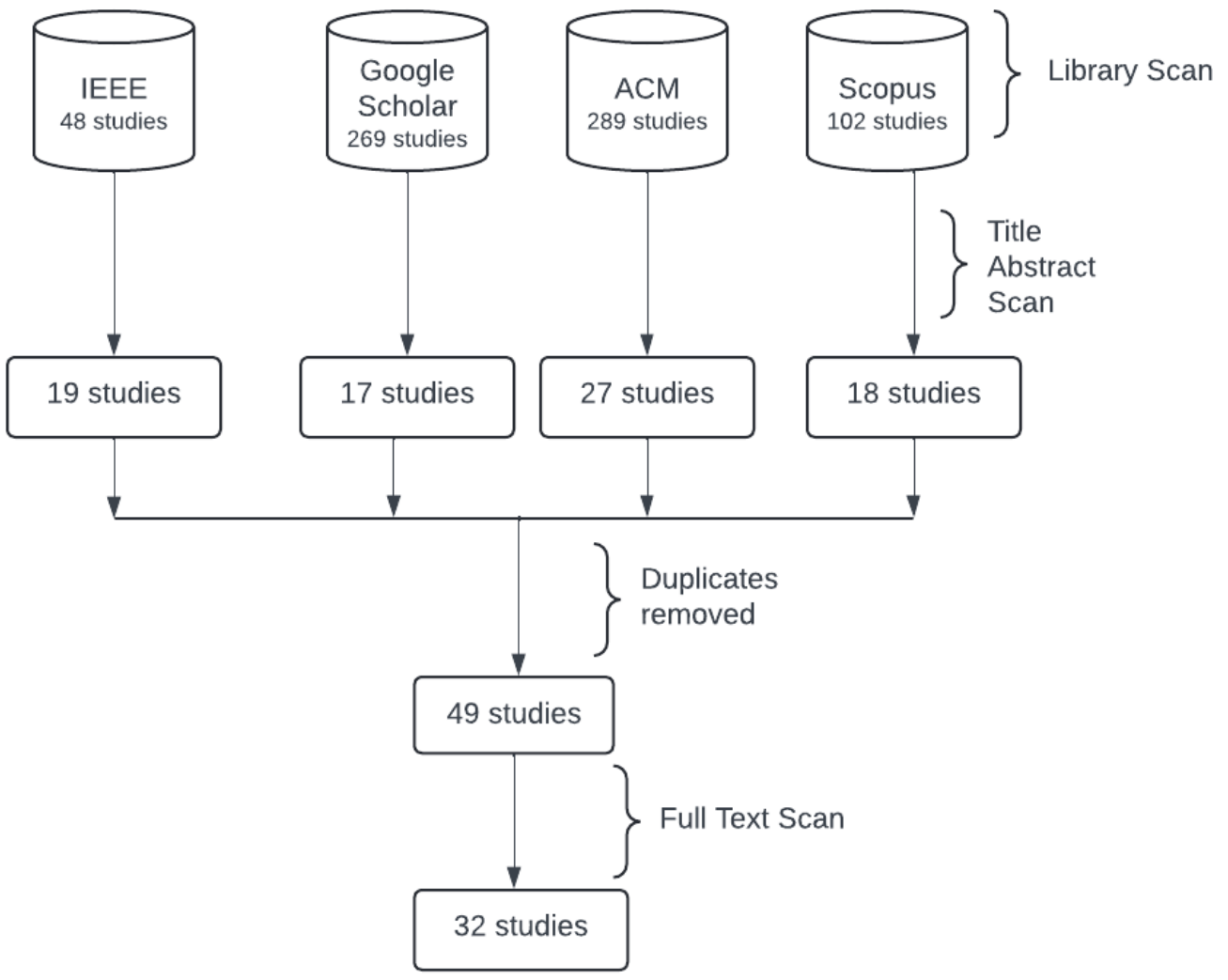

Fig 2. Search Process

## B. Data Extraction

Based on the full text scan performed in the previous step, 32 papers were examined in this study and categorized into three MLOps pipelines in the table 3, which has been deductively chosen from the [11]. While reviewing the studies, we also observed some relevant research on sustainable ML pipelines, which covers the explainability and sustainability of the former three obtained MLOps pipelines. The following information has been mapped as various pipelines in Table 1.

TABLE 1 Metadata of Included Studies.

| Study References | Year | MLOps Pipelines |
|---|---|---|
| 43 | 2023 | MD |
| 44 | 2017 | DM |
| 45 | 2020 | Sustainability |
| 46 | 2021 | DM, MD |
| 47 | 2022 | MD |
| 48 | 2023 | DM, MD |
| 49 | 2022 | Sustainability |
| 50 | 2022 | DM, MC, MD |
| 51 | 2022 | MD |
| 52 | 2023 | DM |
| 53 | 2024 | MC |
| 54 | 2022 | MD, MC |
| 55 | 2022 | MC |
| 56 | 2021 | MD |
| 57 | 2022 | MD |
| 58 | 2023 | MD |
| 59 | 2024 | DM, MD |
| 60 | 2023 | DM |
| 61 | 2024 | Sustainability |
| 62 | 2022 | DM, MD |
| 63 | 2021 | MD |
| 64 | 2021 | MD |
| 65 | 2023 | DM, MD |
| 66 | 2023 | MD |
| 67 | 2024 | MD |
| 68 | 2022 | DM, MD |
| 69 | 2020 | MD |
| 70 | 2021 | MD |
| 71 | 2022 | DM, MD |
| 72 | 2023 | DM, MD |
| 73 | 2023 | MD |
| 74 | 2022 | MD |

## C. Data Aggregation

After identifying and labeling the papers, we then aggregated counts for the number of publications per year over our timeframe, as shown in Fig. 3.

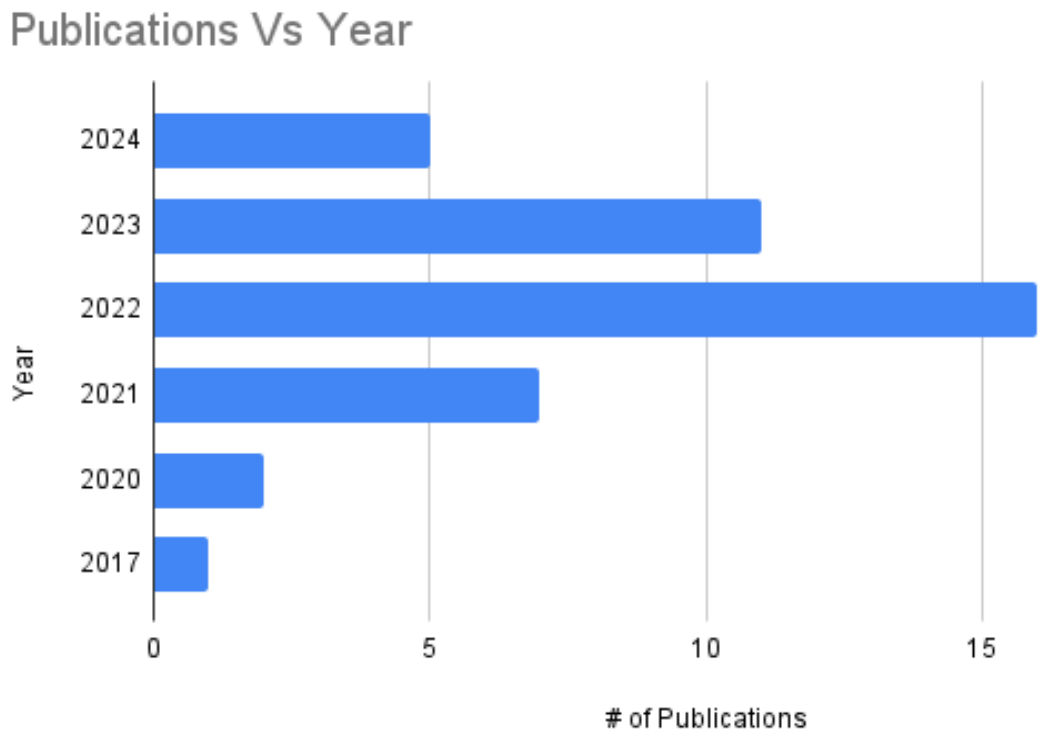

Fig 3 Publications vs. Year

Based on the above chart it is evident that most of the research has been performed during the 2022 and 2023, as depicted by figure 3. This can be possible either due to the availability of new modelling methods or any novelty being introduced among the research community. It is also interesting to observe during this phase the most important pipelines there are being studied are the Model deployment and data management pipelines.

## IV. Results and Discussion

This section provides an overview of the results obtained from the SMS, including a detailed analysis of the quantitative data for each Research Question.

- RQ1:What are the research trends in MLOps, how many studies cover these and how do they converge to different MLOps pipelines?

The data shown in table 2 and fig. 4 demonstrate a strong and growing interest among the MLOps community which depicts that the field of MLOps has been an important field of research among the community. Specifically it is observed that the research trends has been in the model deployment (MD) category in MLOps. There is significant effort focused on addressing the issues of operationalizing machine learning models in production. This is evident and also a confirmation that it has been increased due to the raising interest of MLOps integration in various field of works like manufacturing, mining, marketing etc. Most of the scientific literature has mostly concentrated on the Model Deployment category, namely on Model monitoring, Managing deployment pipelines, and Operations and feedback loops. These are important for integrating a model to the existing production system, as unlike a software deployment where the user can test and perform necessary changes to optimize the new code as per production need prior to merging to the production, in machine learning it is highly unlike to completely test and optimize the model hyper parameters like performance, biasness, accuracy etc. of the model being getting deployed, as the data used for this purpose is limited and controlled (being collected based on specific trends or patters), whereas in production we have various patterns of data which the model has to go through and learn and optimize itself. Hence monitoring of a productional model is an important step of deployment to provide necessary inputs through the feedback loop so the model can learn and improve its hyper parameters. This is clearly apparent, since a significant proportion of the literature, namely over 37%, was only dedicated to this particular problem. While model creation and data management are the other two pipelines which are completed prior to model deployment, they together without model deployment pipelines, cannot ensure a machine learning system is functioning as intended, especially in a dynamic environment like production. Since these are the steps that are performed during initial requirement analysis and model creation and most of the time the teams involved in these stages of pipelines are working in siloed environments and has minimum to no interactions, hence, continuous real-time monitoring of system activity, together with automatic reaction, is essential for ensuring long-term system stability [5].

TABLE 2: Research trends in different MLOps pipelines

| MLOps Pipelines | Research Trends | Study References |
|---|---|---|
| Data Management | Data access and management | [48,50,62,65,72] |
| Data Management | Shortage of diverse data samples | [50,52,71] |
| Data Management | Data cleaning and validation | [44,48,50,51,60] |
| Data Management | Data labeling | [44,50,59,68] |
| Model Creation | Feature Selection | [50,74] |
| Model Creation | Calculations of performance metrics | [50,53,54,55,71] |
| Model Creation | Algorithm & hyper-parameter selection | [50,55,74] |
| Model Creation | Model evaluation | [50,54,55,59,62,71] |
| Model Creation | Experiment tracking | [46,50,55,59] |
| Model Deployment | Model monitoring | [46,47,48,50,51,54,57,63,66,68,69] |
| Model Deployment | Managing deployment pipelines | [46,50,54,56,58,64,65,66,70,71,72,73] |
| Model Deployment | Operations and feedback loops | [43,48,50,51,58,67,69,71,74] |
| Model Deployment | Incompatibilities between dev & prod | [50,51,54,72] |
| Sustainability | Complexity as infrastructure grows | [45,62] |

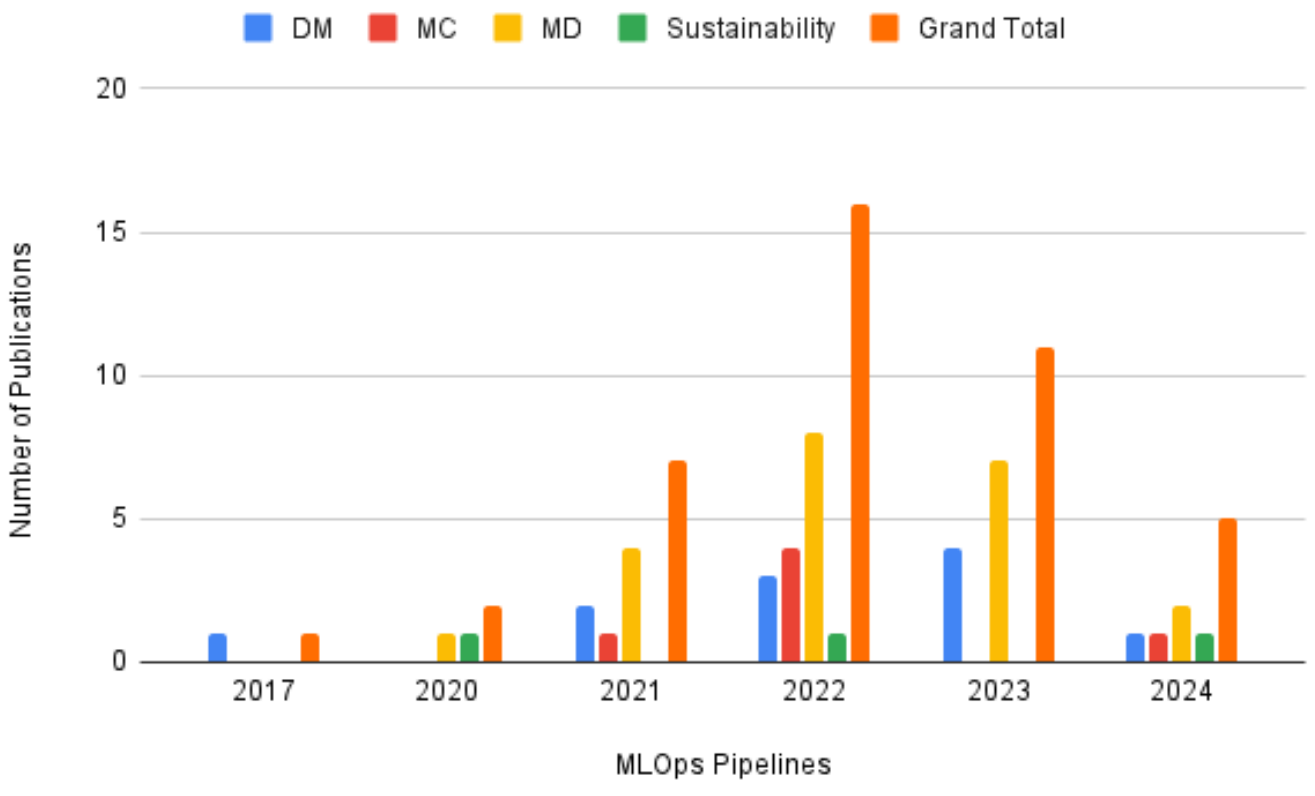

Fig 4: Number of Studies per Year based on MLOps pipelines.

Although Data Management and Model Creation are acknowledged as fundamental, the comparatively low occurrence of studies studying these topics shows they may not be as urgent in contemporary MLOps practice and difficulties or in early phases of research development, even though they are still important. Nevertheless, their existence in the study corpus indicates a comprehensive comprehension that MLOps encompasses more than just deploying machine learning models in production. It also includes model building and data management.

There has been a noticeable increase in both the publishing output and the demand for research in several fields in 2022. This may include the emergence of novel machine learning technology or a reaction to industry need for enhanced operational procedures in machine learning implementations. Based on the summarization Model Deployment (MD) has a predominant interest in the field of MLOps among academic community with proliferating research articles year over year. It comprises of 52% of the studies reviewed in this work. The next in the queue is Data Management (DM) which occupies 26% of the studies, it is evident with advent rise in the big data technologies and computer vision, implementation of ML models for obtaining tangible outcomes has been an area of interest and hence some challenges are mostly being faced in these areas.

The study of combined research findings from various years reveals that the field is actively striving to find a balance between the flexibility of machine learning model creation and the speed of agile software engineering. The many methodologies used by researchers demonstrate a research community that is flexible and quick to address the difficulties posed by machine learning settings. Although Model Deployment is the main focus of the study, the discipline acknowledges the significance of a holistic approach that encompasses Data Management and Model Creation, despite the fact that these areas need greater representation in the existing body of literature. This indicates a research community that is attentive to the current needs of MLOps and foresees its future direction.

- RQ2: What kind of novelty or new ideas do these studies constitute in relation to the pipelines?

The solutions and tools mentioned to these research trends from all the sources are aggregated and categorized in the table. These solutions help us give an insight on how these are being proposed for utilization for MLOps operationalization overcome in the industry and academia.

TABLE 3. MLOPS PIPELINES RESEARCH TRENDS AND NOVELTY MAPPING

| MLOps Pipelines | Research Trends | Novelty & Studies |
|---|---|---|
| Data Management | Data access and management | Datalake [50,62,72] |
| Data Management | Shortage of diverse data samples | Resampling, Augmentation [52] |
| Data Management | Data cleaning and validation | Data scrubbing [48,51] |
| Data Management | Data labeling | Use trained model to label [50] |
| Model Creation | Feature Selection | Feature store [53,55] |
| Model Creation | Calculations of performance metrics | River [54] |
| Model Creation | Algorithm & hyper-parameter selection | AutoML [11] |
| Model Creation | Model evaluation | Deepchecks [55,71] |
| Model Creation | Experiment tracking | DVC [46] |
| Model Deployment | Model monitoring | Model picker, Evidently AI [54,68,69] |
| Model Deployment | Managing deployment pipelines | Efficient automated pipelines right from development [46,50] |
| Model Deployment | Operations and feedback loops | Kubeflow [43] |
| Model Deployment | Incompatibilities between dev & prod | Conduct feasibility study [54] |

The table depicts the growing interest in MLOps and specifically in the areas of Data management and Model deployment as critical parts outside the standalone model creation pipeline work. Though it is a need to pass on the feedback to the model creation pipeline for continuous training and testing, still it is clear that the operationalization is mostly influenced by the model deployment and data availability for the team working on it to progress. They affect machine learning project efficiency, scalability, and reliability: Effective data intake pipelines merge data from several sources into a single repository. Data validation, cleansing, and enrichment which are part of data management are essential for model reliability, integrity, security, and compliance where rules are maintained via data governance policies.

Versioning datasets track changes and ensure experiment repeatability. Preprocessing raw data into features for model training is crucial.

Model health depends on monitoring performance and reporting forecasts, mistakes, and other indicators. Versioning models allow tracking, comparing, and rolling back iterations.

Streamlined data management and Automated model deployment pipelines speed up model development and production. Effective data management and model deployment allow ML operations to scale to bigger datasets and inference loads. Scalable architectures allow models to forecast in real-time for many users. Reliable data management and deployment pipelines decrease failures and maintain performance. Reliable systems boost ML confidence and decrease production downtime.

Data management and model deployment are essential to MLOps, improving machine learning project efficiency, scalability, dependability, reproducibility, and cost.

## V. LIMITATIONS

Every mapping study has limitations, we address ours here.

*External Validity*: This is the generalizability of extrapolating the findings of a scientific study beyond its specific setting. Put simply, it refers to the degree to which the findings of a study may be applied or transferred to different contexts, individuals, stimuli, and time periods. Generalizability pertains to the extent to which a predetermined sample may be applied to a wider population, whereas transportability refers to the extent to which one sample can be applied to another target population. To ensure generalizability of our findings, To establish external validity, we have diligently determined the "scope" of the study, which pertains to the extent to which the theory or argument of this study can be applied or its restrictions. We made a concerted effort to include a diverse range of studies covering various aspects of MLOps. However, it’s important to acknowledge that there are other variations in research trends, and regions might not fully represent global practices.

*Construct Validity*: Construct validity concerns whether the methods and measures accurately reflect the studied phenomena. Our systematic mapping study aimed to provide a comprehensive overview of MLOps research trends. The diverse definitions and practices within MLOps may have led to variations in how it is conceptualized and implemented. Furthermore, researcher bias and reliance on published literature, excluding unpublished or non-peer-reviewed studies, also threaten the comprehensiveness of our findings

*Conclusion Validity*: This threat concerns the ability to draw accurate conclusions from the data. Our study’s conclusions are based on the analysis of 32 studies, which, while comprehensive, might only encompass some relevant research. The potential for publication bias, where only positive or significant results are published, could skew our findings. Furthermore, the evolving nature of MLOps and ways of implementation may need to be revisited in future for inclusion of new research.

*Additional Limitations*: One notable limitation is the potential for selection bias in our study selection process, as the inclusion criteria and databases used may have inadvertently excluded relevant studies. We also recognize that the rapid evolution of both MLOps means that our study may not capture the most recent developments in the field.

## VI. Conclusions

The most effective method for incorporating machine learning models into production is to apply MLOps. Each year, a greater number of businesses use these strategies, and more study is conducted in this field. A fully developed MLOps system that employs continuous training has the potential to lead us to machine learning models that are both more efficient and more realistic. Additionally, selecting the appropriate tools for each individual task is an ongoing issue. Although there are a great number of papers pertaining to the many tools, it is not simple to adhere to the principles and include them in the most effective manner. There are times when we are forced to make a decision between flexibility and resilience, each of which comes with its own set of advantages and disadvantages. Monitoring is the last step, and it is one of the most important points of interest that must be considered. Monitoring the condition of the whole system via the use of sustainability, robustness, fairness, and explainability is, in our opinion, the most important factor in developing mature, automated, robust, and efficient MLOps systems. Considering this, it is of the utmost importance to create models and methods that are capable of enabling this sort of monitoring, such as explainable machine learning models. Future work would be to address in building explainable machine-learning models for production.

## VII. Future Work

There is a considerable research gap in the sustainability and monitoring of MLOps pipelines in the development of comprehensive frameworks and standardized metrics for assessing and benchmarking the environmental impact of machine learning models throughout their lifecycle.

*Standardized Sustainability Metrics*: While there is growing awareness of the need for sustainable AI practices, there is a lack of standardized metrics for measuring the environmental impact of machine learning models. Research is needed to develop comprehensive frameworks that consider factors such as energy consumption, carbon emissions, and resource utilization across different stages of the MLOps pipeline.

*Lifecycle Assessment Tools*: Existing tools for assessing the environmental impact of machine learning models often focus on specific stages of the lifecycle, such as training or inference. There is a need for integrated lifecycle assessment tools that provide a holistic view of the environmental footprint of MLOps pipelines, from data collection to model deployment and decommissioning.

*Real-time Monitoring and Feedback*: Real-time monitoring and feedback mechanisms are essential for detecting inefficiencies and optimizing sustainability in MLOps pipelines. Research is needed to develop automated monitoring tools that provide real-time insights into energy consumption, carbon emissions, and resource utilization, enabling proactive optimization and decision-making.

Addressing these research gaps will contribute to the development of more sustainable and environmentally responsible MLOps pipelines, enabling organizations to minimize their carbon footprint and contribute to a more sustainable future.

## Included Studies